\documentclass[twocolumn]{aastex631}

\begin{document}

\title{SgrA$^{*}$ spin and mass estimates through the detection of an extremely large mass-ratio inspiral}

\author[0000-0002-9458-8815]{Ver\'{o}nica V\'{a}zquez-Aceves}
\affiliation{Institute of Applied Mathematics \\
Academy of Mathematics and Systems Science, Chinese Academy of Sciences \\
Beijing 100190, China}

\author[0000-0003-2509-6558]{Yiren Lin}
\affiliation{Astronomy Department \\
School of Physics, Peking University \\
Beijing 100871, China}

\author[0000-0002-5467-3505]{Alejandro Torres-Orjuela}
\email{Corresponding author: atorreso@mail.sysu.edu.cn}
\affiliation{MOE Key Laboratory of TianQin Mission \\
TianQin Research Center for Gravitational Physics \& School of Physics and Astronomy \\
Frontiers Science Center for TianQin, Gravitational Wave Research Center of CNSA \\
Sun Yat-Sen University (Zhuhai Campus) \\
Zhuhai 519082, China}



\begin{abstract}

Estimating the spin of ${\rm SgrA^\ast}$ is one of the current challenges we face in understanding the center of our Galaxy. In the present work, we show that detecting the gravitational waves (GWs) emitted by a brown dwarf inspiraling around ${\rm SgrA^\ast}$ will allow us to measure the mass and the spin of ${\rm SgrA^\ast}$ with unprecedented accuracy. Such systems are known as extremely large mass-ratio inspirals (XMRIs) and are expected to be abundant and loud sources in our galactic center. We consider XMRIs with a fixed orbital inclination and different spins of ${\rm SgrA^\ast}$ ($s$) between 0.1 and 0.9. For both cases, we obtain the number of circular and eccentric XMRIs expected to be detected by space-borne GW detectors like LISA and TianQin. We find that if the orbit is eccentric, then we expect to always have several XMRIs in band while for almost circular XMRIs, we only expect to have one source in band if ${\rm SgrA^\ast}$ is highly spinning. We later perform a Fisher matrix analysis to show that by detecting a single XMRI the mass of ${\rm SgrA^\ast}$ can be determined with an accuracy of the order $10^{-2}\,\text{M}_{\odot}$, while the spin can be measured with an accuracy between $10^{-7}$ and $10^{-4}$ depending on the orbital parameters of the XMRI.

\end{abstract}

\keywords{black hole physics (159); gravitational waves (678); supermassive black holes (1663); Milky Way galaxy physics (1056)}


\section{Introduction}

${\rm SgrA^\ast}$, the supermassive black hole (SMBH) at our galactic center, was recently observed by the Event Horizon Telescope~\citep{SgrA1_2022, SgrA2_2022, SgrA3_2022, SgrA4_2022, SgrA5_2022, SgrA6_2022} obtaining an estimated mass of $4\times10^6\,\text{M}_{\odot}$, which is in good agreement with previous estimates~\citep{Ghez_2008,Gillessen_2009}; in contrast, measuring its spin remains a major challenge. In this work we show that detecting a single extremely large mass ratio inspiral (XMRI), i.e., a  brown dwarf (BD) inspiraling towards ${\rm SgrA^\ast}$ due to energy loss by gravitational waves (GWs), is enough to determine the spin and mass of ${\rm SgrA^\ast}$ with very good accuracy.

When an XMRI forms in our galactic center, its GW emission can be detected by space-borne detectors such as the Laser Interferometer Space Antenna (LISA)~\citep{LISA_2017,Pau_2020} and TianQin~\citep{TianQin_2016, TianQin_2021}. Furthermore, its large mass-ratio $q= m_{\rm BD}/ M_{\rm SgrA^\ast} \approx 10^{-8}$, allows the BD to spend a large amount of time inspiraling around the SMBH, and the slow evolution of the orbit simplifies the analysis of its gravitational radiation. Understanding its formation channels and evolution is a key part of the description that will lead to accurate templates to identify and extract the information encoded within its gravitational radiation.

In a dense stellar system, such as our galactic center, two-body relaxation processes slowly change the orbits of the orbiting objects by diffusion in energy and angular momentum ($J$). Diffusion in $J$ is more efficient than diffusion in energy, so the eccentricity ($e$) of an orbit changes faster than its semimajor axis ($a$). Consequently, the orbit's pericentre (R$_{\rm p}$) is perturbed, allowing objects to reach very close distances to the central black hole. An inspiraling system is formed when a compact object is diffused into an orbit with a small R$_{\rm p}$ such that after just one pericentre passage, the orbit evolves only due to the energy lost by gravitational radiation.

Once such an inspiraling system forms, it emits GWs that can potentially be detected by space-based detectors such as LISA and TianQin. Considering ${\rm SgrA^\ast}$ as the bigger component of such a source was first proposed in \citet{freitag_2003} where main-sequence starts (MSs) orbiting ${\rm SgrA^\ast}$ were studied. In \citet{Barack:2003fp}, MSs around ${\rm SgrA^\ast}$ are also studied but going to masses as low as $0.06\,\text{M}_\odot$. The mass is comparable to the mass of a BD, and just as low-mass  MSs, BDs are convective and can be described by a polytropic equation~\citep{Burrows_1993}; however, BDs are predominantly composed of metallic hydrogen and helium supported by electron degeneracy, as they can not sustain hydrogen fusion. MSs are more easily disrupted, so \citet{Barack:2003fp} restrict the analysis to a source with a pericenter distance of more than ten Schwarzschild radii, which significantly diminishes the accuracy obtained. In \citet{gourgoulhon_le-tiec_2019}, a variety of compact objects orbiting ${\rm SgrA^\ast}$, including BDs are studied. However, they restricted their studies to circular equatorial orbits and, as we will show later, XMRIs are most likely to form at highly eccentric orbits with moderate inclinations. ${\rm SgrA^\ast}$ as a component of a GW source was also studied in \citet{berry_gair_2013}, where extreme mass-ratio bursts are considered. Although the rate for such bursts from our galactic center is comparable to that of XMRIs, their short-lived nature considerably reduces the information we can obtain about ${\rm SgrA^\ast}$ in comparison to long-lasting sources like XMRIs.

Recent estimates show that at the time the LISA and TianQin missions will be in space, there could be about 15 eccentric and five circular XMRIs in our galactic center emitting GWs in a detectable range~\citep{Pau_2019}. In Sec.~\ref{sec:nox}, we compute the number of circular ($e\lesssim$ 0.2) and eccentric ($e\gtrsim$ 0.9) sources that will be present in the LISA/TianQin band in the cases in which the spin ($s$) of ${\rm SgrA^\ast}$ takes a value of 0.1 and 0.9. The XMRIs considered in the present work are composed of BDs inspiraling in prograde orbits with a fixed orbital inclination of $i=0.1\,\text{rad}$. In Sec.~\ref{sec:msm}, we perform a Fisher matrix analysis of the systems described in Sec.~\ref{sec:nox} and show, for each spin case, how accurately we can measure the mass and spin of ${\rm SgrA^\ast}$ by detecting an individual XMRI either in the eccentric or in the circular regime. We draw conclusions and give an outlook to other questions about ${\rm SgrA^\ast}$ that might be addressed using XMRIs in Sec.~\ref{sec:con}

\section{Number of XMRIs in the center of our galaxy}\label{sec:nox}

To estimate the number of XMRIs ($N_{\rm tot}$) in the center of our galaxy, we multiply the event rates ($\dot{\Gamma}_{\rm XMRI}$) by the merger timescale ($T_{\rm GW}$) of a typical source. The event rates  are estimated as in \citet{Pau_2019} by integrating those sources of which the orbits evolve only due to GW emission in phase space. Such systems are defined by a semimajor axis $a \lesssim a_{\rm crit}$, where $a_{\rm crit}$ marks the critical distance at which the system reaches an internal energy that shields the binary against relaxation processes; for XMRIs $a_{\rm crit}$ is of the order of $10^{-3}\,{\rm pc}$ to $10^{-4}\,{\rm pc}$ and depends on the spin magnitude of the SMBH and the orbital inclination. The rates are also sensitive to the model used to describe the density distribution $\rho(r)$ of BDs and BHs around Sgr A*. Observations and numerical simulations have shown that stellar objects around an SMBH distribute in a cusp; \cite{BW} predicted that main sequence stars form a cusp around the SMBH of the form $\rho(r)\approx r^{-\gamma}$, where $\gamma=1.75$ for stellar mass black holes and $\rho(r)\approx r^{-\beta}$, $\beta=1.5$ for the less massive objects such as main sequence stars. By performing numerical simulations \cite{Marc_2006} finds that the value of $\beta$ falls between 1.3 and 1.4, while in strong mass segregation scenarios in which the number of heavy objects is relatively low and barely interact with each other, suggest $\gamma\in (2,2.75)$ and $\beta \in (1.5, 1.75)$ \citep{AlexanderTal_2009}. However, the extensive analysis performed by \cite{Gallego-Cano_2018, Schodel_2018, Baum_2018}, in which they combine observations and N-body simulations of the Milky Way's nuclear star cluster, showed that stars form a cusp described by $\beta \in (1.13 -1.43)$ suggesting that strong mass segregation does not occur. In the present work, we assume that BDs follow the same distribution as main-sequence stars and that stellar-mass black holes lead the relaxation processes; to compute the number of sources, we take the conventional Bahcall $\&$ Wolf values $\gamma=1.75$ and $\beta=1.5$ and although it has been shown that strong mass segregation can enhance the event rates of extreme mass-ratio inspirals \citep{Preto_2010, Amaro-Seoane_2011}, which is relevant for galaxies with similar characteristics as our Galaxy, we do not consider the strong mass segregation scenario as we focus on the center of the Milky Way.

The XMRIs event rate is given by

\begin{equation}\label{eq:EventRate}
    \dot{\Gamma}_{\rm XMRI} = \frac{3}{5 T_0}\frac{ N_{\rm IR}}{\text{R}_{\rm h}^{\lambda}} f_{\rm sub}\left\lbrace\left[ a_{\rm crit}^{\lambda}\left(\text{ln}\left(\frac{a_{\rm crit}}{2 \text{R}_{\rm L}}\right)- \frac{1}{\lambda}\right)\right] \right\rbrace,
\end{equation}
where $\lambda=9/2 - \beta - \gamma$, $N_{\rm IR} = M_{\rm  SgrA^\ast} / \bar{m}$ is the number of objects within the influence radius of ${\rm SgrA^\ast}$, R$_{\rm h}\approx 3\,\text{pc}$, $T_0\approx 6.9\times10^9\,\text{yr}$ is a normalization timescale, $\bar{m} = 0.27\,\text{M}_{\odot}$ is the average stellar mass, $f_{\rm sub}=0.21$ is the fraction number of BDs obtained from a Kroupa broken power law~\citep{Kroupa_2001}, and R$_{\rm L}$ is the maximum value between the position of the last stable orbit (LSO), and the tidal disruption radius $r_{\rm tidal}$, obtained as

\begin{equation}\label{eq:rtidal}
    r_{\rm tidal}= \left( 2 \frac{(5-n) M_{\rm  SgrA^\ast}}{3 m_{\rm BD}}\right)^{1/3} r_{\rm BD},
\end{equation}
where $r_{\rm BD}$ is the radius of a BD described as a polytrope with a polytropic index $n=1.5$~~\citep{shapiro_teukolsky_1983}. The position of the LSO is shifted depending on the spin magnitude; an object orbiting around a Kerr black hole in a prograde orbit finds that the LSO is shifted towards the Schwarzschild radius so it can reach closer distances to the SMBH. Consequently, an XMRI formed around an SMBH with high spin can reach smaller pericentre distances and higher eccentricities than an XMRI formed around a slowly spinning SMBH. The position of the last stable orbit of a Kerr black hole is given by R$_{\rm LSO}$ = 4 R$_{\rm S} \mathcal{W}(i,s)$, where the effect of the spin and inclination with respect to the spin axis is included through the function $\mathcal{W}(i, s)$~\citep{Pau_2013} and R$_{\rm S}=2GM/c^2$ is the Schwarzschild radius, $G$ is the gravitational constant, and $c$ is the speed of light in vacuum. \\

Considering a BD with a mass $m_{\rm BD}=0.05\,{\rm M_{\odot}}$, and a radius $r_{\rm BD}=0.083\,{\rm R_{\odot}}$~\citep{Sorahana_2013} we obtain $r_{\rm tidal}\approx 2.8\,{\rm R_S}$, which is inside the LSO for a Schwarzschild SMBH; however, for a highly spinning SMBH with $s=0.9$, the position of the LSO shifts to $r_{\rm LSO}\approx 1.7\,{\rm R_S}$ and, hence, we consider $r_{\rm tidal}$ as the limit to compute the event rates.

An XMRI orbits ${\rm SgrA^\ast}$ for around $10^8$ years; its merger timescale ($T_{\rm GW}$) is given by
\begin{equation}\label{eq:mergertimescale}
    T_{\rm GW}= \frac{10}{64}\frac{a^4\, c^5}{G^3 \,m\, M^2} f(e),
\label{eq:TGW}
\end{equation}
where
\begin{equation}
    f(e)=(1-e^2)^{7/2}\left[1+\frac{73}{24}e^2+ \frac{37}{96} e^4\right]^{-1}\, ,
\end{equation}
with $m=m_{\rm BD}$, and $M=M_{\rm SgrA^\ast}$. Note that Equation~(\ref{eq:mergertimescale}) applies to systems with $M \gg m$ and, where the pericentre remains approximately constant during the inspiral process, which is suitable for cases where $e \approx 1$ as we are considering here. The commonly used Peters' timescale~\citep{Peters:1964} expressed as
\begin{equation}
    T_{\rm P} = \frac{5}{256} \frac{a^4 c^5}{G^3 M m (M+m)} f(e)
    \label{eq:Peterstimescale}
\end{equation}
assumes that the secular evolution of the eccentricity can be neglected, leading to inaccurate results when applied to highly eccentric and relativistic systems; therefore, Peters' timescale expressed as in Eq.~(\ref{eq:Peterstimescale}) must not be implemented to obtain the merger timescale of inspiraling systems. In \cite{Zwick_2020} and \citep{Zwick_2021}, a set of correction factors, named $R$, $Q_{\rm h}$ and $Q_{\rm s}$ are presented. By multiplying Eq.~(\ref{eq:Peterstimescale}) by the factor $R$, Peters' timescale is corrected for secular eccentricity evolution, and when multiplied by $Q_{\rm h}$ and $Q_{\rm s}$, 1.5 Post-Newtonian (PN) hereditary fluxes and spin-orbit couplings are added. These correction factors do not significantly affect the event rates \citep{VVA_2022}; however, they can slightly change the number of sources in the detection band; therefore, for comparative purposes, we estimate the number of sources taking the merger time scale $T_{\rm GW}$ (Equation~\ref{eq:TGW}) as well as the corrected time scale $T_{\rm RQ}=T_{\rm P} \times R Q_{\rm h}Q_{\rm s}$ as the merger timescale. The explicit formulas of the correction factors $R$, $Q_{\rm h}$ and $Q_{\rm s}$, can be found in \citep{Zwick_2021}.

The total number of XMRIs $N_{\rm tot}$ orbiting around ${\rm SgrA^\ast}$ estimated as
\begin{equation}\label{eq:N_tot}
    N_{\rm tot}= \dot{\Gamma}_{\rm XMRI} (a_{\rm crit})  \times  T_{\rm GW},
\end{equation}
does not take into account if the GWs emitted by the system are within the detection band of space-borne detectors. To obtain the number of sources within the detection band, we follow the procedure presented in \citet{Pau_2019} where a continuity function to describe the evolution of the sources in the phase space is defined. With this, it is possible to estimate the number of XMRIs at a given point of their evolution, allowing us to estimate the number of eccentric ($N_{\rm II}$) and circular ($N_{\rm I}$) XMRIs formed with an initial eccentricity $e_0$. $N_{\rm II}$ is the sum of all the sources from the moment they start to emit GWs in the LISA/TianQin detection band; at this point, the sources are still highly eccentric. As the orbit evolves, it reaches a point where the second harmonic is dominant, marking the beginning of the circular regime; these sources can reach a signal-to-noise ratio (SNR) of $\gtrsim2,000$, while for the eccentric regime, the SNR is around 30.

To perform our analysis on the spin and mass of ${\rm SgrA^\ast}$, we focus on two representative spin values: $s=0.1$ and $s=0.9$ and compute the orbital parameters of XMRIs formed around an SMBH of $4\times 10^6 M_{\odot}$ by BDs of $0.05\,\text{M}_\odot$ in prograde orbits with an orbital inclination of $i=0.1\,\text{rad}$, which maximizes the value of the event rates.

For each value of the semimajor axis $a\lesssim a_{crit}$, there is a range of eccentricities, typically high ($e>0.9$), at which an XMRI can form as long as $e \lesssim e_{\rm max}$, where
\begin{equation}\label{eq:emax}
    e_{\rm max}=1-\frac{\text{R}_{\rm L}}{a};
\end{equation}
if a compact object approaches an SMBH in an orbit with $e>e_{\rm max}$, it crosses the LSO (or the disruption radius) at pericentre and suffers a direct plunge (or a tidal disruption). In the considered scenario, the initial eccentricity of XMRIs ranges between 0.99 and $\gtrsim$0.999 for a semimajor axis between $4\times 10^{-4}$pc and $8\times 10^{-3}$pc. We obtain a set of XMRIs with orbital parameters within this range for each spin value and compute the number of sources that are emitting GWs inside the LISA and TianQin detection band. In Table~\ref{tab:Nsources}, we show the average number of sources ($\bar{N}$) obtained for the two representative spin values, where $N_{\rm RQ}$ represents the number of sources obtained by taking the corrected merger timescale $T_{\rm RQ}$ instead of $T_{\rm GW}$ (Equation~\ref{eq:TGW}) as the merger timescale; we obtain an average of $\bar{N}_{\rm II}=8^{+9}_{-3}$ eccentric sources for the slowly spinning case and $12^{+6}_{-4}$ for the highly spinning case, where we also find one circular source. When computing $\bar{N}_{\rm RQ_I}$ and $\bar{N}_{\rm RQ_{II}}$, we obtain a similar number of sources. Note that for the cases of no source or only one source, we do not give a range because the difference from the main value is smaller than one but we only can have an integer number of sources.

\begin{table}\center \renewcommand{\arraystretch}{1.35}
\begin{tabular}{|c|c|c|c|c|} \hline
SgrA$^{*}$ spin &$\bar{N}_{\rm I}$& $\bar{N}_{\rm II}$ &$\bar{N}_{\rm RQ_I}$& $\bar{N}_{\rm RQ_{II}}$ \\ 
\hline 
 0.1  & 0 & 8$^{+9}_{-3}$ & 0 & 9$^{+10}_{-4}$  \\ 
\hline
 0.9  & 1 & 12$^{+6}_{-4}$ & 1 & 11$^{+5}_{-3}$  \\ 
\hline
\end{tabular} 
\caption{Average number of circular ($\bar{N}_{\rm I}$) and eccentric ($\bar{N}_{\rm II}$) XMRIs in band. The subscript $RQ$ indicates the rates calculated with the corrected Peters' time scale instead of the timescale $T_{\rm GW}$ (Equation~\ref{eq:TGW}).}
\label{tab:Nsources}
\end{table} 

In the next section, we study how accurately we can measure the mass and spin of ${\rm SgrA^\ast}$ by detecting a single XMRI, which represents a lower limit according to our estimates for $\bar{N}_{\rm I}$ and $\bar{N}_{\rm II}$ as well as for previous results by~\cite{Pau_2019}. A detailed study on the number of sources in the LISA/TianQin band with different inclinations, and the accuracy we can achieve on the measurements of the mass and spin by detecting multiple XMRIs, will be published by the current authors soon elsewhere.

\section{Measuring the spin and mass of S\lowercase{gr}A$^\ast$}\label{sec:msm}

As mentioned in the previous section, the XMRIs we study have a SNR well above the typical detection threshold for comparable sources~\citep{babak_gair_2017,fan_hu_2020}. Therefore, we estimate how accurately the spin and the mass of ${\rm SgrA^\ast}$ can be measured by performing a Fisher matrix analysis~\citep{coe_2009}, which provides a linearized estimate for the measurement errors that asymptotes to the true errors in the high-SNR limit. Having the waveform of the XMRI $h(\mathbf{\theta})$, where $\mathbf{\theta}$ are the intrinsic and extrinsic parameters of the source, the Fisher matrix can be defined as
\begin{equation}\label{eq:deffm}
    \Gamma_{i,j} := \left\langle\frac{\partial h(\mathbf{\theta})}{\partial\theta_i},\frac{\partial h(\mathbf{\theta})}{\partial\theta_j}\right\rangle.
\end{equation}
Here $\langle\cdot,\cdot\rangle$ represents the noise-weighted inner product~\citep{finn_1992,klein_barausse_2016}. The inverse of the Fisher matrix $C := \Gamma^{-1}$ then approximates the sample covariance matrix of the Bayesian posterior distribution for the parameters given the observed signals.

A more detailed study of the measurability of the spin would involve analyses such as in \citet{cutler_vallisneri_2007}, full posterior sampling~\citep{li_2013} or analyses based on convolutional neural networks~\citep{zhang_messenger_2022}. However, we only intend to estimate the accuracy with which the spin of the SMBH in the center of our galaxy can be measured. Moreover, considering the very high SNR of the XMRIs, the Fisher matrix analysis provides very good estimates, and thus, we stick to the analysis introduced above.

We generate the waveform for the XMRIs using a code based on \citet{Barack:2003fp} and \citet{Fang_Huang_2020} expanding the rate of change of the orbital frequency and eccentricity to 2.5 post-Newtonian (PN) order while considering the effect of the spin of the SMBH through the pericentre precession and the Lense-Thirring precession to effectively 1.5 PN order. Restricting our analysis to lower PN orders reduces the computational costs and allows us to include a higher number of modes which is crucial due to the high eccentricity of the orbit and the sensitivity of space-based detectors~\citep{LISA_2017,TianQin_2016}.

Recent results from the Event Horizon Telescope~\citep{SgrA1_2022, SgrA2_2022, SgrA3_2022, SgrA4_2022, SgrA5_2022, SgrA6_2022} indicate that the spin of ${\rm SgrA^\ast}$ is pointing at a small angle ($\le \pi/6\,\text{rad}$) towards Earth. Therefore, we consider the case where the spin of ${\rm SgrA^\ast}$ is aligned with the line-of-sight for different spin magnitudes varying between the two cases introduced in `Number of XMRIs in the center of our galaxy': a low spin of 0.1 and a high spin of 0.9. For all cases, we consider an almost circular XMRI with an eccentricity $e = 0.21$ and a pericentre distance R$_{\rm p}$ = 4.33 R$_{\rm S}$ as well as a highly eccentric XMRI with an eccentricity $e = 0.84$ and a pericentre distance R$_{\rm p}$ = 7.42 R$_{\rm S}$. For the more circular case, we consider the first ten modes, while for the more eccentric one, we consider a total of 50 modes ($n=27 - 76$) which are the modes that have the highest SNR in the detection band of LISA. We set ${\rm SgrA^\ast}$ to have a mass of $4\times10^6\,\text{M}_\odot$ and to be located at a distance of $8\,\text{kpc}$ from Earth, and, for both cases, we set the mass of the BD to be $0.05\,\text{M}_\odot$, consider its orbit to have an inclination of $0.1\,\text{rad}$, and an observation time of two years.

\begin{figure}
\includegraphics[width=0.5\textwidth]{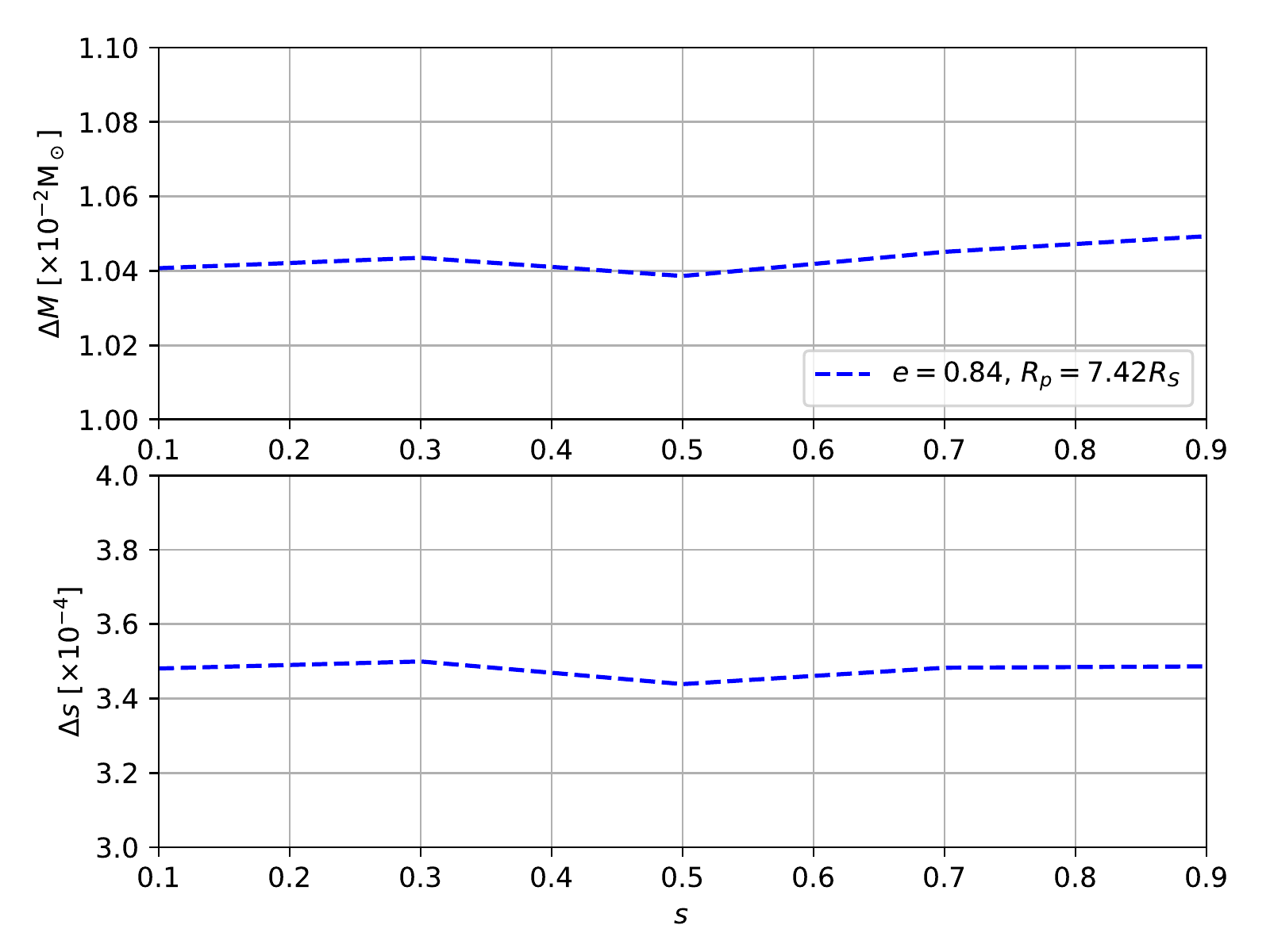}
\caption{
  The accuracy for the measurement of the mass and the spin of ${\rm SgrA^\ast}$ for different spin of ${\rm SgrA^\ast}$ in the case of a highly eccentric XMRI ($e=0.84$, R$_{\rm p}$ = 7.42 R$_{\rm S}$).
}\label{fig:ecce}
\end{figure}

\begin{figure}
\includegraphics[width=0.5\textwidth]{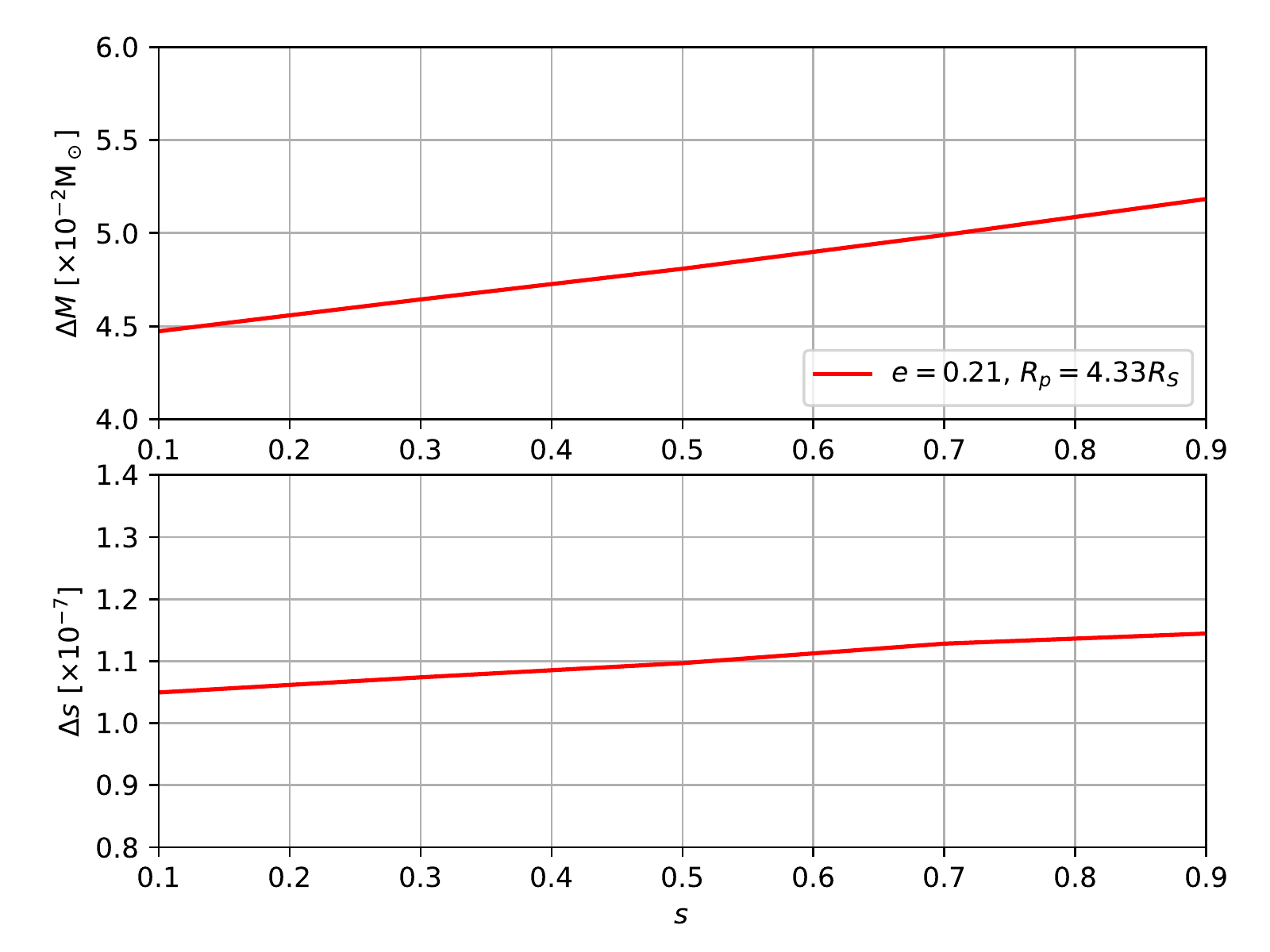}
\caption{
  The accuracy for the measurement of the mass and the spin of ${\rm SgrA^\ast}$ for different spin of ${\rm SgrA^\ast}$ in the case of an almost circular XMRI ($e=0.21$, R$_{\rm p}$ = 4.33 R$_{\rm S}$).
}\label{fig:circ}
\end{figure}

Figure~\ref{fig:ecce} shows the accuracy for the measurement of the mass and spin of ${\rm SgrA^\ast}$ with a highly eccentric XMRI. We see that the mass is measured with an accuracy of the order $10^{-2}\,\text{M}_\odot$ while the spin is measured with an accuracy of the order $10^{-4}$. The accuracy obtained varies slightly for different spin values, but this difference is most likely a numerical feature rather than a physical effect. In particular, for the relative error in the spin $\delta s := \Delta s/s$ we get an increased accuracy for an increasing spin where $\delta s$ is of the order of $10^{-3}$ for $s=0.1$ and of the order of $10^{-4}$ for $s=0.9$. The accuracy for the measurement of the mass and the spin of ${\rm SgrA^\ast}$ from an XMRI with an almost circular orbit is shown in Figure~\ref{fig:circ}. In this case, the accuracy in the measurement of the mass is also of the order of $10^{-2}\,\text{M}_\odot$, again with a small dependence on the spin that, however, is not significant and probably just a result of numerical inaccuracies. In contrast, for the spin, the accuracy increases to an order of $10^{-7}$, again with a dependence on the spin that is not very significant. Moreover, the relative error $\delta s$ of the spin accuracy is again decreasing for an increasing spin ($\sim10^{-6}$ for $s=0.1$ and $\sim10^{-7}$ for $s=0.9$) as expected.

Note that the accuracy of the mass seems to decrease for circular XMRIs while the accuracy for the spin increases significantly. However, for both types of orbit, the accuracy for the mass of ${\rm SgrA^\ast}$ is of the order of $10^{-2}\,\text{M}_\odot$, and the difference can be attributed to the inaccuracy of the Fisher matrix analysis rather than to physical features. The increase in the accuracy for the spin for almost circular orbits, in contrast, can be explained by the fact that this orbit has a smaller radius but still considerable eccentricity, and thus, it is affected stronger by relativistic perihelion precession and Lense-Thirring precession. These two effects that depend on the spin of ${\rm SgrA^\ast}$ are also the reason why the relative error for the spin decreases when the spin of ${\rm SgrA^\ast}$ increases. Therefore, we conclude that the mass of ${\rm SgrA^\ast}$ will be measured with an accuracy of $10^{-2}\,\text{M}_\odot$ for different orbits while its spin will be measured with an accuracy of $10^{-4}$ for highly eccentric orbits and with an accuracy of $10^{-7}$ for almost circular orbits.

\section{Conclussions}\label{sec:con}

XMRIs have recently been proposed as a new source for space-borne detectors such as LISA or TianQin~\citep{Pau_2019}. Due to the small mass of the secondary compact object (the BD) these sources will only be detectable in the close universe. However, we expect them to be abundant and, most likely, the loudest sources in our galaxy reaching easily a SNR of over 1000. Considering a fixed orbital inclination of $i=0.1\,\text{rad}$, we estimate the number of XMRIs that space-borne detectors will be able to detect when launched, finding that if the spin of ${\rm SgrA^\ast}$ is $s=0.9$, we expect at least one circular XMRI and an order of ten eccentric XMRIs. In case the spin of ${\rm SgrA^\ast}$ is $s=0.1$, we only find eccentric sources. However, these estimates depend highly on the orbital parameters with which the XMRI is formed, so the chance of having an XMRI in the circular regime for low spin cases cannot be ruled out.

Using a waveform model based on \citet{Barack:2003fp} and considering multiple modes of the GW, we perform a Fisher matrix analysis to estimate how accurately the mass and the spin of ${\rm SgrA^\ast}$ can be measured from the detection of one XMRI. We find that the mass will be measured with an accuracy of $10^{-2}\,\text{M}_{\odot}$ for different orbital parameters, and the spin will be measured with an accuracy of the order $10^{-4}$ for highly eccentric orbits while it will be measured with an accuracy of the order $10^{-7}$ for almost circular orbits since these orbits are also closer to merger. As expected, the relative accuracy in the detection of the spin of ${\rm SgrA^\ast}$ will be better for higher spin values, but the absolute error is of the same order independent of the spin.

Future space-borne missions will detect XMRIs with different orbital inclinations. Additionally, the population of XMRIs in the LISA/TianQin band is related to ${\rm SgrA^\ast}$'s spin magnitude. As there is currently no restriction on the magnitude of the spin of ${\rm SgrA^\ast}$, in a follow-up paper that will be published soon elsewhere, we extend this study to different orbital parameters, inclinations and spin values to provide an insight into the accuracy with which we can measure the mass and the spin of ${\rm SgrA^\ast}$. Furthermore, we highlight that in this work, we focus on the mass and spin of ${\rm SgrA^\ast}$ because they are the two most important parameters of a SMBH. However, the astonishing accuracy that can be achieved detecting only one XMRI and considering that multiple XMRIs will be in band when LISA/TianQin are launched, shows that XMRIs are very promising sources for detailed studies of ${\rm SgrA^\ast}$ and the fundamental physics of SMBHs.

\section*{Acknowledgment}

We thank Pau Amaro-Seoane for helping us with his expertise on the formation and properties of XMRIs. We further thank Xian Chen for helpful discussions. VVA acknowledges support from CAS-TWAS President's Ph.D. Fellowship Program of the Chinese Academy of Sciences \& The World Academy of Sciences. YL is supported by the National Key Research and Development Program of China (Grant No. 2021YFC2203002). ATO acknowledges support from the Guangdong Major Project of Basic and Applied Basic Research (Grant No. 2019B030302001) and the China Postdoctoral Science Foundation (Grant No. 2022M723676).


\bibliography{xmrisbib}{}
\bibliographystyle{aasjournal}



\end{document}